\documentclass[preprint,onecolumn,nofootinbib]{revtex4}
\usepackage[colorlinks=true,linkcolor=blue,urlcolor=blue,filecolor=black,citecolor=red,pdfstartview=FitV,pdftitle={},pdfsubject={},pdfkeywords={},pdfpagemode=None,bookmarksopen=true]{hyperref}
\usepackage{graphicx}
\usepackage{amsmath}
\usepackage{amsfonts}
\usepackage{amssymb,ulem}
\usepackage{color,xcolor}%
\usepackage{CJK}
\usepackage{subfigure}
\usepackage{amsthm,amsmath,amssymb}
\usepackage{mathrsfs}
\usepackage{multirow}

\usepackage{dcolumn}
\usepackage{float}
\begin{document}
\title{Quasinormal modes and ringdown waveforms of the Frolov black hole}
\author{Zhijun Song $^{1}$}
\thanks{1873285349@qq.com}
\author{Huajie Gong $^{1}$}
\thanks{huajiegong@qq.com}
\author{Hai-Li Li $^{2}$}
\thanks{1329750467@qq.com, corresponding author}
\author{Guoyang Fu $^{3}$}
\thanks{FuguoyangEDU@163.com}
\author{Li-Gang Zhu $^{1}$}
\thanks{zlgoupao@163.com}
\author{Jian-Pin Wu $^{1}$}
\thanks{jianpinwu@yzu.edu.cn, corresponding author}

\affiliation{
	$^{1}$ \mbox{Center for Gravitation and Cosmology, College of Physical Science and Technology,} \mbox{Yangzhou University, Yangzhou 225009, China}\\
	$^{2}$ \mbox{Basic Teaching Department, Shenyang Institute of Engineering,} \mbox{Shenyang 110136, China}\\
	$^{3}$ \mbox{Department of Physics and Astronomy, Shanghai Jiao Tong University,} \mbox{Shanghai 200240, China}
}

\begin{abstract}

In this paper we investigate scalar perturbation over a Frolov black hole (BH), which is a regular BH induced by the quantum gravity effect. The quasinormal frequencies of a scalar field always consistently reside in the lower half-plane, and the time-domain evolution of the field demonstrates a decaying behavior, with the late-time tail exhibiting a power-law pattern. These observations collectively suggest the stability of a Frolov BH against scalar perturbation. Additionally, our study reveals that the quantum gravity effect leads to slower decay modes. For the case of the angular quantum number $l=0$, the oscillation exhibits non-monotonic behavior with the quantum gravity parameter $\alpha_0$. However, once $l\geq 1$, the angular quantum number surpasses the influence of the quantum gravity effect.

\end{abstract}

\maketitle

\section{Introduction}\label{sec-intro}

Regular black holes (BHs) were originally introduced as a means to circumvent the central singularity inherent in ordinary BHs. Regular BHs may be categorized into two types based on their asymptotic behavior approaching the center: those with a de-Sitter (dS) core and those with a Minkowskian core. Notable examples of regular BHs with a dS core are the Bardeen BH \cite{Bardeen:1968}, the Hayward BH \cite{Hayward:2005gi}, and the Frolov BH \cite{Frolov:2016pav}. A regular BH with a Minkowskian core is usually characterized by the exponential potentials, as described in references \cite{Xiang:2013sza,Culetu:2013fsa,Culetu:2014lca,Rodrigues:2015ayd,Simpson:2019mud,Ghosh:2014pba,Ghosh:2018bxg,Li:2016yfd,Martinis:2010zk,Ling:2021olm}. Comprehensive reviews on regular BHs can be found in references \cite{Bambi:2023try,Vagnozzi:2022moj,Lan:2023cvz}. This research aims to analyze the characteristics of the quasinormal modes (QNMs) of the Frolov BHs.

Perturbing a BH and observing its response is widely recognized as a powerful method for extracting crucial characteristics of the BH. This perturbation can be implemented either by introducing a probe matter field into its spacetime by hand or by physically perturbing its metric. Perturbing the metric leads to the emission of gravitational waves (GWs). Prior to reaching equilibrium, the system undergoes a phase of BH merging, referred to as the ringdown phase. During this stage, the BH releases GWs with characteristic discrete frequencies, i.e. quasinormal frequencies (QNFs). These frequencies encode information about the decaying scales and damped oscillations of the BH \cite{Berti:2009kk}. Studying the properties of QNFs offers an opportunity to detect deviations from General Relativity (GR) or even quantum gravity effects through observations of GWs. However, the majority of regular BHs are typically constructed by incorporating quantum gravity effects at the phenomenological level, making it challenging to establish consistently effective gravitational perturbation equations. Fortunately, even when only a probe matter field is considered over these regular BHs, their QNM spectra are also influenced by the background spacetime. Therefore, these QNM spectra also provide crucial information about the internal structure of the BH and can be used to model the GW during the ringdown phase at a phenomenological level \cite{Berti:2005ys,Berti:2018vdi,Fu:2022cul,Fu:2023drp,Gong:2023ghh}.

In \cite{Lopez:2018aec}, the authors investigated the properties of QNMs for a probe massless scalar field over a Frolov BH in the eikonal limit, i.e. a large angular quantum number. The effective potential for a Frolov BH is found to be higher than that of the uncharged Hayward BH. The imaginary parts of the QNFs for a Frolov BH grow with the charge and exhibit a maximum, followed by a more pronounced decrease for small values of the parameter associated with the effective cosmological constant at small distances. As the charge grows so do the real parts of the QNFs. In this work, we will conduct a systematic investigation of the QNMs of Frolov BH, with a primary focus on the case of low angular quantum numbers, in contrast to the eikonal limit studied in \cite{Lopez:2018aec}.

Our paper is organized as follows. In section \ref{sec1}, we offer a concise overview of Frolov BHs along with their fundamental properties, and we also introduce the dynamics of a scalar field over a Frolov BH. In section \ref{secPSmethod}, we provide a concise introduction to the pseudospectral method. The characteristics of the QNMs and ringdown waveforms of the scalar field over a Frolov BH are discussed in section \ref{sec2} and \ref{sec3}, respectively. Conclusions and further discussions are presented in Section \ref{sec4}.

\section{Probe scalar field over a Frolov black hole}\label{sec1}

\begin{figure}[ht]
	\centering{
		\includegraphics[width=5.2cm]{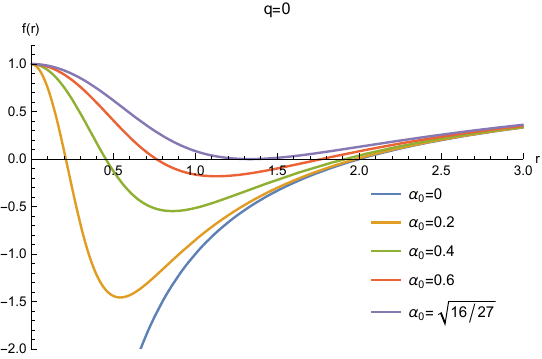}\hspace{0.2cm}
		\includegraphics[width=5.2cm]{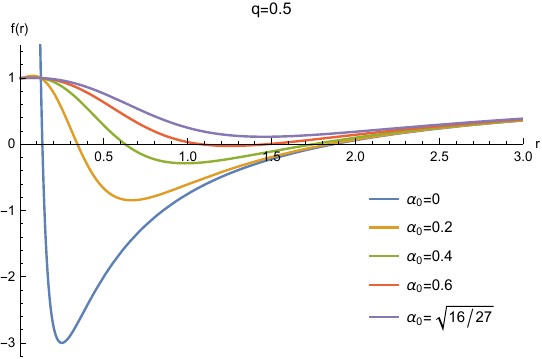}\hspace{0.2cm}
		\includegraphics[width=5.2cm]{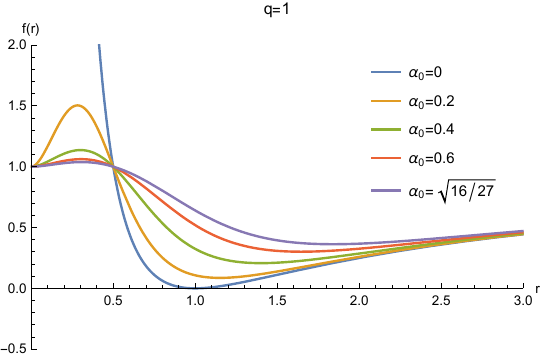}\
		\caption{The metric function $f(r)$ of a Frolov BH for various $q$ and $\alpha_0$.}
		\label{fvr}
	}
\end{figure}

\begin{figure}[ht]
	\centering{
		\includegraphics[width=8cm]{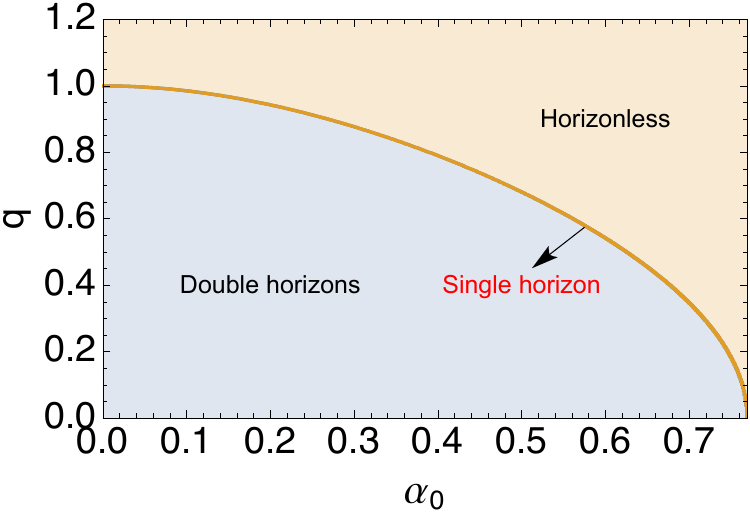}
		\caption{The phase diagram $\alpha_0$ versus $q$, illustrating the horizon structure of Frolov spacetime.}
		\label{Horizon}
	}
\end{figure}

A Frolov BH is an extension of a Hayward BH to the charged case, originally proposed in \cite{Frolov:2016pav}. The geometry of a Frolov regular BH is described by \cite{Frolov:2016pav}:
\begin{eqnarray}
	&&
	\textrm{d}s^2=-f(r)\textrm{d}t^2+\frac{1}{f(r)}\textrm{d}r^2+r^2\textrm{d}\theta^2+r^2\sin^2\theta\textrm{d}\phi^2\,,
	\label{Frolov-metric-v1}
	\
	\\
	&&
   f(r)=1-\frac{(2Mr-q^2)r^2}{r^4+(2Mr+q^2)\alpha_0^2}\,,
	\label{Frolov-metric-v2}
\end{eqnarray}
where $M$ denotes the BH mass. This core of a Frolov BH is characterized by an effective cosmological constant $\Lambda=3/\alpha_0^2$, where $\alpha_0$ represents the Hubble length. The Hubble length characterizes a universal hair and is bounded by the following inequality \cite{Hayward:2005gi}:
\begin{eqnarray}
	\alpha_0\leq \sqrt{16/27}M\,.
	\label{Hayward-constrain}
\end{eqnarray}
Satisfying this constraint leads to significant quantum gravity effects. Subsequently, we will set $M=1$ for simplicity, without loss of generality. The charge parameter $q$ characterizes a specific hair and satisfies $0\leq q\leq 1$. At $q=0$, a Frolov BH reduces to a Hayward BH, and at $\alpha_0=0$, it simplifies to the Reissner-Nordström (RN) BH. It is obvious that when both $q=0$ and $\alpha_0=0$, a Frolov BH reduces to a Schwarzschild BH. Figure \ref{fvr} shows the metric function $f(r)$ of a Frolov BH for various $q$ and $\alpha_0$. We begin by considering the case of $q=0$, corresponding to the Hayward BH (the left plot in Figure \ref{Hayward-constrain}). As $\alpha_0$ increases from the Schwarzschild BH case, the Hayward BH evolves into a BH with double horizons. On further increasing $\alpha_0$ to the upper bound defined by the inequality \eqref{Hayward-constrain}, the Hayward BH transforms into a BH with a single horizon. When $q \neq 0$, with an increasing $\alpha_0$, the Frolov geometry, initially featuring double horizons, transitions to a spacetime with a single horizon (the middle plot in Figure \ref{fvr}). Further increments in $\alpha_0$ lead to the development of a horizonless spacetime, often describing the exotic compact objects (the middle plot in Figure \ref{fvr}). However, when $q$ is increased to $q=1$, the Frolov geometry describes a horizonless spacetime for all $\alpha_0>0$ (the right plot in Figure \ref{fvr}). Figure \ref{Horizon} displays the phase diagram of $\alpha_0$ versus $q$, illustrating the horizon structure of Frolov spacetime. In this paper, we will exclusively focus on studying the case of BH spacetime with horizons.

\begin{figure}[ht]
	\centering{
		\includegraphics[width=8cm]{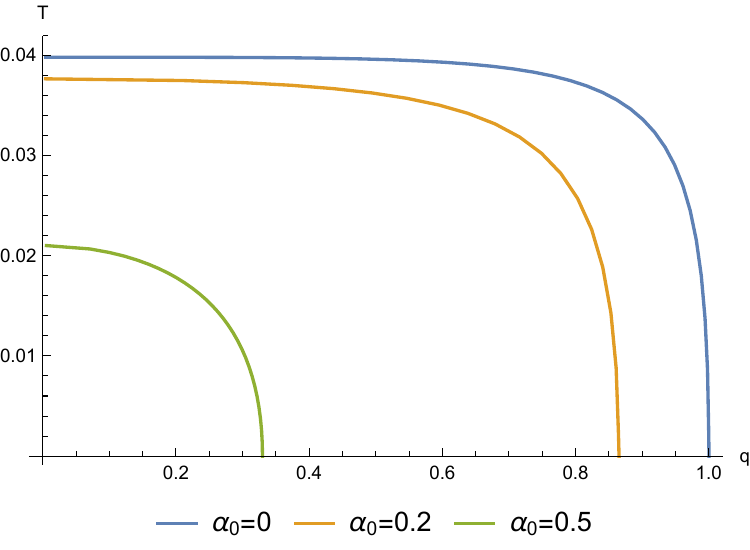}
		\caption{The Hawking temperature as a function of $q$ with different $\alpha_{0}$.}
		\label{Temvsq}
	}
\end{figure}

The Hawking temperature of the regular BH can be worked out as:
\begin{eqnarray}
	T=\frac{f'(r_h)}{4\pi}=\frac{r_h \left(-q^2 \left(2 \alpha _0^2
		r_h+r_h^4\right)-4 \alpha _0^2
		r_h^2+r_h^5+\alpha _0^2 q^4\right)}{2 \pi 
		\left(2 \alpha _0^2 r_h+r_h^4+\alpha _0^2
		q^2\right){}^2}.
\end{eqnarray}
Here, $r_h$ represents the event horizon of the BH. From the equation above, it is evident that when $q=0$ and $T=0$, the upper limit of equation \eqref{Hayward-constrain} is satisfied. Figure \ref{Temvsq} illustrates the relationship between the Hawking temperature and the charge parameter $q$ for various values of $\alpha_0$ to aid visualizing.

The dynamics of the probe scalar field can be described by the following Klein-Gordon (KG) equation:
\begin{eqnarray}
	\frac{1}{\sqrt{-g}}\partial_\nu(g^{\mu\nu}\sqrt{-g}\partial_\mu\Phi)=0\,.
	\label{scalar_eq}
\end{eqnarray}
Given the spherical symmetry of the geometry under investigation, we can employ the following spherical harmonics to separate the variables:
\begin{eqnarray}\label{separate}
	\Phi(t,r,\theta,\phi)=\sum_{l,m}Y_{l,m}(\theta,\phi)\frac{\Psi_{l,m}(t,r)}{r}\,,
\end{eqnarray}
In the above equation, $Y_{l,m}(\theta,\phi)$ is the spherical harmonics. Here, $l$ and $m$ stand for the angular and azimuthal quantum numbers, respectively. For a given $l$ and $m$, we have simplified the notation by denoting $\Psi_{l,m}(t,r)$ as $\Psi$. Then, the KG equation \eqref{scalar_eq} may be reformulated in the following manner:
	\begin{eqnarray}\label{Sch_like_eq}
		-\frac{\partial^2\Psi}{\partial t^2}+\frac{\partial^2\Psi}{\partial r_{\ast}^2}   - V_{\text{eff}}\Psi = 0\,,
	\end{eqnarray}
where $r_*$ is the tortoise coordinate associated with $r$, as $\textrm{d}r_{*}/\textrm{d}r=1/f(r)$. The effective potential is given by:
\begin{eqnarray}\label{V_eff1}
	V_{\text{eff}}=f(r)\frac{l(l+1)}{r^2}+\frac{f(r)f'(r)}{r}\,,
\end{eqnarray}
with $l=0, 1, \ldots$.

\begin{figure}[ht]
	\centering{
		\includegraphics[width=5.2cm]{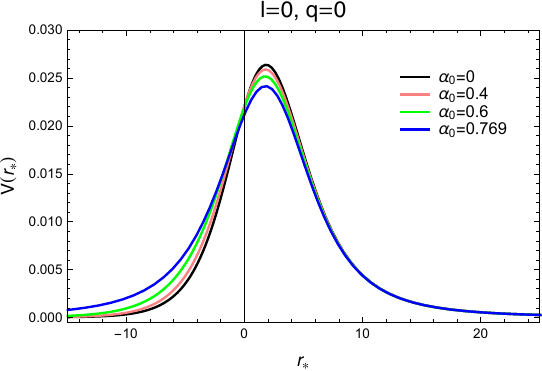}\hspace{0.2cm}
		\includegraphics[width=5.2cm]{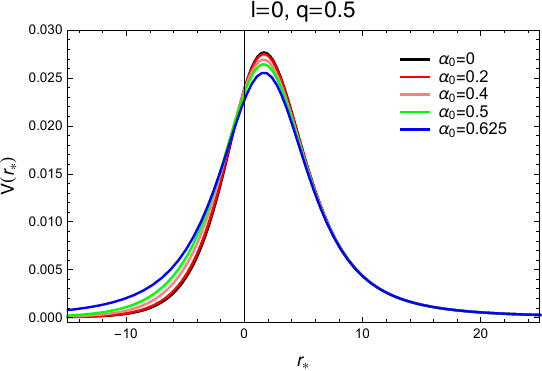}\hspace{0.2cm}
		\includegraphics[width=5.2cm]{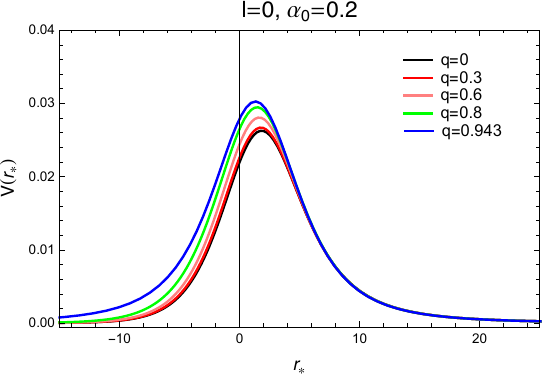}\
		\caption{The effective potential $V(r_*)$ for various BH parameter $\alpha_0$ and charge parameter $q$ with $l=0$.}
		\label{Vr}
	}
\end{figure}
Figure  \ref{Vr} depicts the effective potential $V(r_*)$ for different BH parameters, namely $\alpha_0$ and $q$, with $l=0$. For a fixed charge $q$, it is observed that as $\alpha_0$ increases, the peak value of the effective potential decreases, whereas the position of the peak remains almost unchanged (the left and middle plots in Figure \ref{Vr}).
The right-most plot in Figure \ref{Vr} shows the case of changing $q$ with fixed $\alpha_0=0.2$. With an increase in $q$, both the peak value of the effective potential and its position consistently shift to the left.
Moreover, it is noteworthy that, regardless of whether we vary $\alpha_0$ or $q$, there are discernible changes in the near-horizon behavior. These characteristics of the effective potential undoubtedly influence the behaviors of the QNMs.

\section{Pseudospectral Method}\label{secPSmethod}
Numerous techniques have been devised for determining QNMs, with the pseudospectral approach being one of the potent numerical instruments. This method has been successfully applied in various models to compute QNFs (see, e.g. \cite{Jansen:2017oag,Wu:2018vlj,Fu:2018yqx,Xiong:2021cth,Liu:2021fzr,Liu:2021zmi,Jaramillo:2020tuu,Jaramillo:2021tmt,Destounis:2021lum,Fu:2022cul,Fu:2023drp,Gong:2023ghh}).

We will work in the Eddington–Finkelstein coordinate system and our study will be carried out in the frequency domain. To this end, we employ the following transformations:
\begin{eqnarray}
	r\to 1/u \ \ \text{and} \ \   \Psi=\textrm{e}^{\textrm{i} \omega r_*(u)}\psi\,.
	\label{transv1}
\end{eqnarray}
Then we need to impose the ingoing and outgoing boundary condition at the horizon and infinity, respectively,
\begin{equation}
	\Psi \sim \textrm{e}^{\mp \textrm{i} \omega r_*}\,,~~~~~~ \;r_*\rightarrow \mp \infty\,.
	\label{BC-infinity}
\end{equation}
Collecting the equations above, the wave equation \eqref{Sch_like_eq} is then transformed into
\begin{eqnarray}
		&&
		(-l(l+1)u^2-2iu(1-2(1+u)f(u)+u(1+2u)f'(u))\omega+4(1+2u)(1-f(u)
		\nonumber
		\\
		&&
		-2uf(u))\omega^2)\psi(u)+(u^4f'(u)+2iu^2(1-(2+4u)f(u))\omega)\psi'(u)+u^4 f(u) \psi''(u)=0\,,
		\label{waveEQ-v2}
\end{eqnarray}
where the prime represents the derivative with respect to $u$. After the transformation \eqref{transv1}, the metric function $f(u)$ becomes
\begin{equation}
	f(u)=1+\frac{u(-2+q^2 u)}{1+2u^3 \alpha_{0}^2+q^2 u^4 \alpha_{0}^2}\,.
	\label{fv2}
\end{equation}

Using the pseudospectral method to solve the eigenvalue problem in $\omega$, the pivotal step involves discretizing the wave equation \eqref{waveEQ-v2} (For more details, please refer to \cite{Jansen:2017oag}).
To achieve this, we employ the Chebyshev grids and Lagrange cardinal functions, defined as follows:
\begin{eqnarray}
	\label{Cg-Lcf}
	u_i=\cos\left( \frac{i}{N}\pi\right) \,, \ \ C_j(u)=\prod_{i=0,i\neq j}^N \frac{u-u_i}{u_j-u_i}\,, \ i=0\,, ...\,, N\,.
\end{eqnarray}
Then, the function $\psi(u)$ can be approximated as
\begin{equation}
	\psi(u)\approx\sum_{j=0}^Nf(u_j)C_j(u)\mathrm,
\end{equation}
where the cardinal functions are linear combinations of Chebyshev polynomials $T_n(u)$ of the first kind:
\begin{equation}
	C_j(u)=\frac2{Np_j}\sum_{m=0}^N\frac1{p_m}T_m(u_j)T_m(u)\mathrm{~,~}p_0=p_N=2\mathrm{~,~}p_j=1\mathrm{~.}
\end{equation}
Afterward, we can obtain the generalized eigenvalue equation in the form
\begin{eqnarray}\label{eq1}
	(M_0+\omega M_1)\psi=0,
\end{eqnarray}
where $M_i$ ($i=0,1$) represents the linear combination of the derivative matrices $D_{ij}^{(n)}=C_i^{n}(u_j)$, where $n$ represents the $n$th derivative matrix. By solving the eigenvalue function, we can determine the QNFs.

\section{Quasinormal modes}\label{sec2}

\begin{figure}[ht]
	\centering{
		\includegraphics[width=8cm]{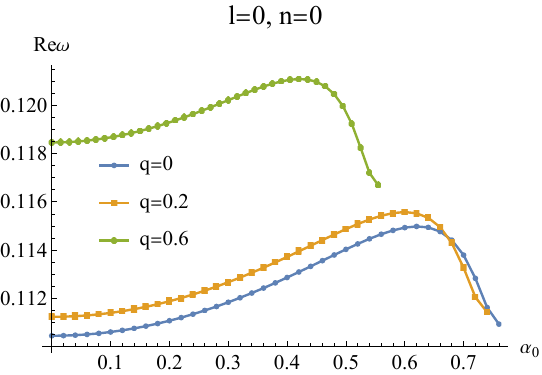}\hspace{2mm}
		\includegraphics[width=8cm]{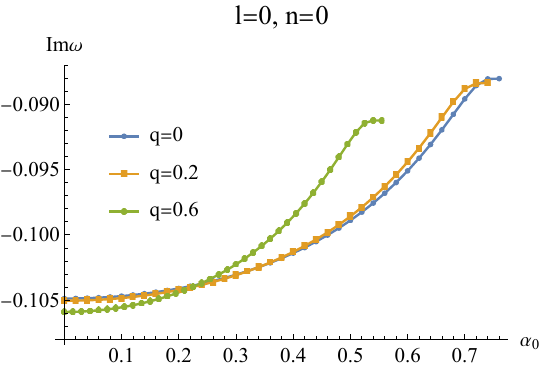}\ \\
		\caption{QNFs as a function of the BH parameter $\alpha_0$ for different charge parameter $q$ with $l=0$ and $n=0$.}
		\label{QNMs-l0}
	}
\end{figure}
\begin{figure}[ht]
	\centering{
		\includegraphics[width=8cm]{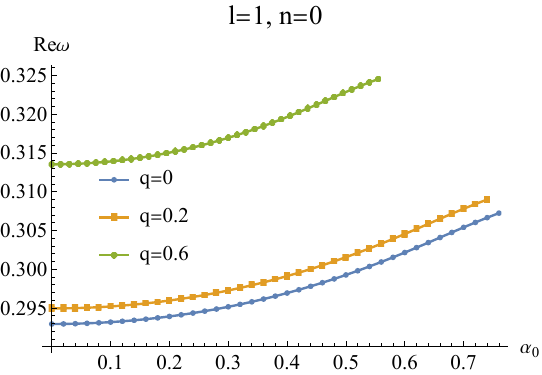}\hspace{2mm}
		\includegraphics[width=8cm]{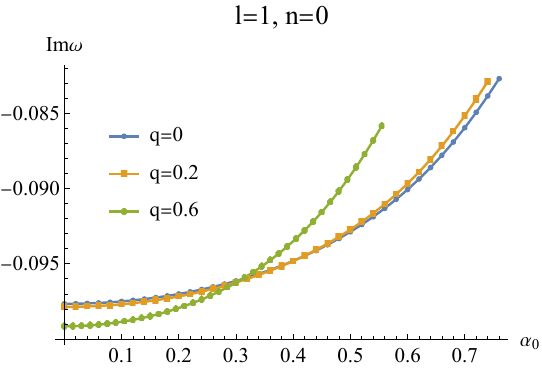}
		\caption{QNFs as a function of the BH parameter $\alpha_0$ for different charge parameter $q$ with $l=1$ and $n=0$.}
		\label{QNMs-l1}
	}
\end{figure}

\begin{table}[ht!]
	\begin{tabular}{c|ccc}
		\hline ${q}$ & $\omega\left(\alpha_{0}= 0\right)$ &  $\omega\left(\alpha_{0}= 1/10\right)$ &  $\omega\left(\alpha_{0}= 1/2\right)$   \\
		\hline	$0	$& 0.11046 - 0.10490\textrm{i}  & 0.11061 - 0.10471\textrm{i} &  0.11403 - 0.09889\textrm{i}  \\
		\hline	$2/10$& 0.11124 - 0.10506\textrm{i} & 0.11140 - 0.10486\textrm{i} & 0.11487 - 0.09858\textrm{i} \\
		\hline	$6/10$& 0.11846 - 0.10593\textrm{i} & 0.11867 - 0.10557\textrm{i} & 0.11974 - 0.09271\textrm{i}
	\end{tabular}
	\caption{The fundamental modes for different BH parameter $\alpha_0$ and charge parameter $q$ with $l=0$.}
	\label{table_l0}
\end{table}
\begin{table}[ht!]
	\begin{tabular}{c|ccc}
		\hline ${q}$ & $\omega\left(\alpha_{0}= 0\right)$ &  $\omega\left(\alpha_{0}= 1/10\right)$ &  $\omega\left(\alpha_{0}= 1/2\right)$   \\
		\hline	$0	$& 0.29294 - 0.09766\textrm{i} & 0.29318 - 0.09750\textrm{i} &  0.29926 - 0.09287\textrm{i}  \\
		\hline	$2/10$& 0.29494 - 0.09786\textrm{i} & 0.29520 - 0.09769\textrm{i} & 0.30155 - 0.09273\textrm{i}   \\
		\hline	$6/10$& 0.31353 - 0.09915\textrm{i} & 0.31390 - 0.09886\textrm{i} & 0.32286 - 0.08913\textrm{i}
	\end{tabular}
	\caption{The fundamental modes for different BH parameter $\alpha_0$ and charge parameter $q$ with the $l=1$.}
	\label{table_l1}
\end{table}

In this section, we will explore the characteristics of the QNMs on a Frolov BH.
We illustrate the behavior of the fundamental modes in relation to the BH parameter $\alpha_0$ for various charge parameters $q$ with $l=0$ in Figure \ref{QNMs-l0} and $l=1$ in Figure \ref{QNMs-l1}. Additionally, for illustrative purposes, we present selected values of the fundamental modes corresponding to various BH parameters $\alpha_0$ and charge parameters $q$ with $l=0$ in Table \ref{table_l0} and $l=1$ in Table \ref{table_l1}.
It is observed that the imaginary parts consistently reside in the lower half-plane. This suggests that a Frolov BH remains stable when subjected to scalar perturbations.

Now, we will delve deeper into studying some particular characteristics of the QNFs for the fundamental modes with respect to both the BH parameter $\alpha_0$ and the charge parameter $q$. When $l=0$, there is a clear non-monotonic pattern in the real parts of QNFs with respect to the BH parameter $\alpha_0$ for a given value of $q$ (see the left plot in Figure \ref{QNMs-l0}). The phenomenon of non-monotonic behavior has been documented in loop quantum gravity (LQG)-corrected BH, as reported in \cite{Fu:2023drp,Gong:2023ghh}. To be more precise, with an increase in the parameter $\alpha_0$, signifying that the system is farther from the Schwarzschild BH or a RN BH, the real parts of the QNFs initially go up, which means that the system is oscillating more strongly, and go down, which means that the oscillation is weakening. As the charge parameter $q$ increases, the turning point of this non-monotonic behavior goes down. It is important to note that the real parts are smaller than that of a Schwarzschild BH or a RN BH as the parameter $\alpha_0$ increases. At the same time, as $\alpha_0$ increases so do the values of the imaginary parts of the QNFs, which points to a lower damping rate (see the right plot in Figure \ref{QNMs-l0}). This result suggests that when quantum gravity effects are present, we have a slower decay modes.

Furthermore, if we fix the BH parameter $\alpha_0$, we observe that in the region of small $\alpha_0$, the real parts of QNFs increase as the charge parameter $q$ rises, indicating that the charge enhances the oscillation of the system. As the parameter $\alpha_0$ increases, we observe that the curves with $q=0.2$ and $q=0.6$ intesect (see the left plot in Figure \ref{QNMs-l0}), suggesting an opposite trend in the change with $q$. The opposite change trend with $q$ between small $\alpha_0$ and large $\alpha_0$ is more pronounced in the imaginary parts (see the right plot in Figure \ref{QNMs-l0}). We also observe that when $\alpha_0$ is relatively large, the curves intersect again.

Next, we examine in detail the scenario when $l=1$, as shown in Figure \ref{QNMs-l1}. The non-monotonic behavior ceases to exist at $l=1$. Specifically, both the real and imaginary parts consistently and continuously increase as the parameter $\alpha_0$ decreases. This means that the system oscillates more strongly and decays more slowly. we also analyzed the QNFs for the values of $l>1$ and arrived at a similar result as in the case when $l=1$. An analogous discovery has been documented with the fundamental modes of a LQG-corrected BH \cite{Gong:2023ghh}.  This means that the angular number has a bigger effect on the fundamental modes than quantum gravity effects. Further research should be done in the future into the universality of the non-monotonic behavior for $l=0$ and the reasons behind this phenomenon.

\section{Ringdown waveform}\label{sec3}

In this section, we will investigate the time-domain profiles of the scalar field over the Frolov BH. To do this, we need to solve numerically the time-dependent wave equation \eqref{Sch_like_eq} using the following initial Gaussian wave packet:
\begin{eqnarray}
&&
\Psi(r_*, t=0)=\exp\left[{-\frac{(r_*-a)^2}{2b^2}}\right]\,,
\nonumber
\
\\
&&
\Psi(r_*,t<0)=0\,.
	\label{Gaussian_wavepacket}
\end{eqnarray}
Without loss of generality, we fix the tortoise coordinate at $r_*=5$. Notes that, the different of initial conditions \eqref{Gaussian_wavepacket} can not influence the ringdown waveform. After the initial outburst stage, the ringdown waveform is fully determine by the QNMs, also called the quasinormal ringing stage.
\begin{figure}[ht!]
	\centering{
		\includegraphics[width=5cm]{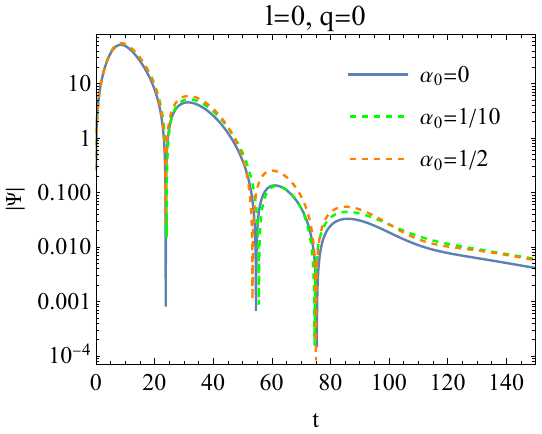}\hspace{1mm}
		\includegraphics[width=5cm]{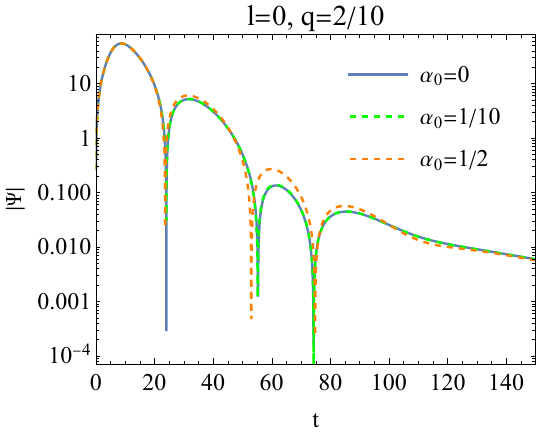}\hspace{1mm}
		\includegraphics[width=5cm]{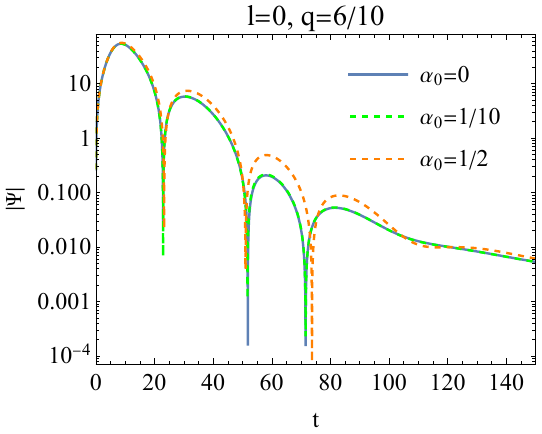}
		\includegraphics[width=5cm]{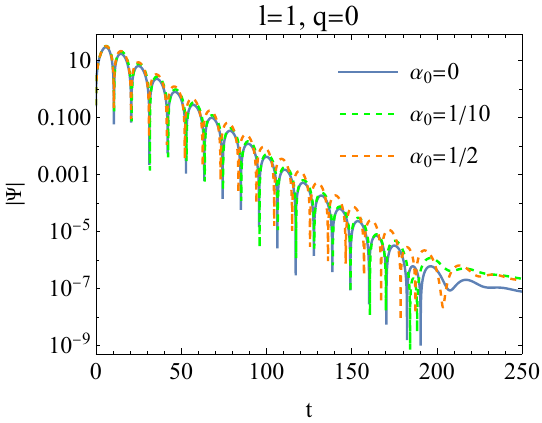}\hspace{1mm}
		\includegraphics[width=5cm]{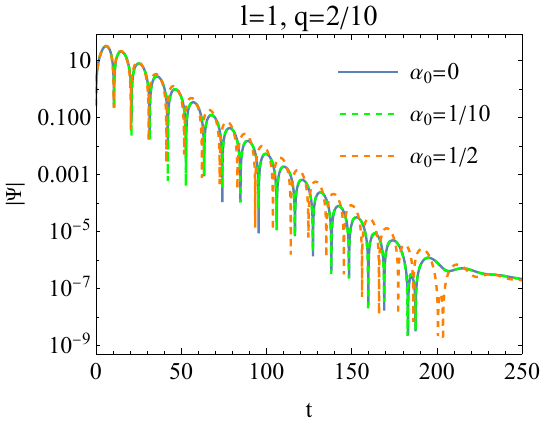}\hspace{1mm}
		\includegraphics[width=5cm]{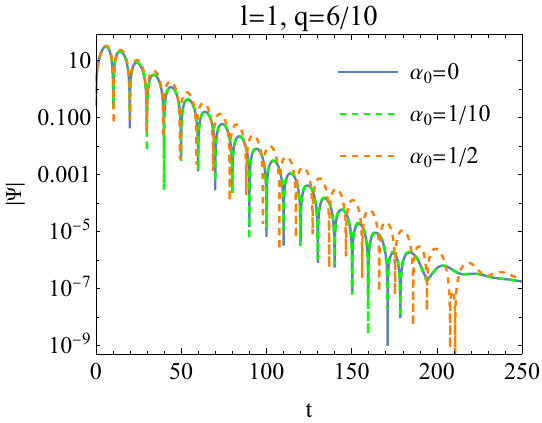}
		\caption{Semilogarithmic plots of the time-domain profiles of the scalar field for $l=0$ and $l=1$  with different parameters of $q$ and $\alpha_0$.}
		\label{TD}
	}
\end{figure}

In Figure \ref{TD} we display the semilogarithmic plots of the time-domain evolution of the scalar field for $l=0$ and $l=1$, showcasing various combinations of $q$ and $\alpha_0$ parameters.
Notably, there is no increase in perturbation over time. Furthermore, we conduct a fitting analysis on the late-time tail decay, revealing a power-law behavior represented as $\Psi \sim t^{-2l+3}$. These findings provide additional confirmation that a Frolov BH is dynamically stable when subjected to perturbations from a massless scalar field. This observation aligns with the findings from QNMs. Additionally, as a validation check, we employ the Prony method to fit the fundamental mode; the results are presented in Tables \ref{table_1} and \ref{table_2}. These findings align with those obtained through the pseudospectral method, as presented in Tables \ref{table_l0} and \ref{table_l1}.

\begin{table}[ht!]
	\begin{tabular}{c|ccc}
	\hline ${q}$ & $\omega\left(\alpha_{0}= 0\right)$ &  $\omega\left(\alpha_{0}= 1/10\right)$ &  $\omega\left(\alpha_{0}= 1/2\right)$   \\
    \hline	$0	$& 0.11089 - 0.10049\textrm{i}  & 0.11401 - 0.096998\textrm{i} &  0.11534 - 0.092569\textrm{i}  \\
	\hline	$2/10$& 0.11462 - 0.09729\textrm{i} & 0.11469 - 0.097132\textrm{i} & 0.11598 - 0.092354\textrm{i} \\
	\hline	$6/10$& 0.11160 - 0.10301\textrm{i} & 0.11781 - 0.098063\textrm{i} & 0.11852 - 0.086906\textrm{i}
	\end{tabular}
\caption{The fundamental modes obtained by the Prony method for different BH parameter $\alpha_0$ and charge parameter $q$ with the $l=0$.}
\label{table_1}
\end{table}
\begin{table}[ht!]
	\begin{tabular}{c|ccc}
		\hline ${q}$ & $\omega\left(\alpha_{0}= 0\right)$ &  $\omega\left(\alpha_{0}= 1/10\right)$ &  $\omega\left(\alpha_{0}= 1/2\right)$   \\
		\hline	$0	$& 0.29287 - 0.097679\textrm{i} & 0.29301 - 0.097522\textrm{i} &  0.29917 - 0.092926\textrm{i}  \\
		\hline	$2/10$& 0.29478 - 0.097902\textrm{i} & 0.29504 - 0.097735\textrm{i} & 0.30146 - 0.092794\textrm{i}   \\
     	\hline	$6/10$& 0.31350 - 0.099311\textrm{i} & 0.31387 - 0.099015\textrm{i} & 0.32288 - 0.089244\textrm{i}
	\end{tabular}
\caption{The fundamental modes obtained by the Prony method for different BH parameter $\alpha_0$ and charge parameter $q$ with the $l=1$.}
	\label{table_2}
\end{table}

\section{Conclusions and discussions}\label{sec4}

In this paper, we have investigated the properties of QNMs of the scalar field over a Frolov BH and have also provided a brief discussion on its time-domain evolution.
A Frolov BH exhibits stability against scalar perturbations, as evidenced by the consistent presence of the imaginary parts of the QNMs in the lower half-plane and the gradual decay of perturbations over time.

When the angular quantum number $l=0$, the real parts of QNFs display distinct non-monotonic behaviors. Specifically, with an increase in the BH parameter $\alpha_0$, the system undergoes stronger oscillations initially, followed by a weakening. The turning point of this non-monotonic behavior shifts downward as the charge parameter $q$ increases. Furthermore, as the parameter $\alpha_0$ increases, the real parts become smaller compared to those of a Schwarzschild BH or RN BH. This suggests that, at this moment, the system undergoes weaker oscillations than a Schwarzschild BH or RN BH. The BH parameter $\alpha_0$ contributes to reducing the damping rate, as indicated by the increase in the values of the imaginary parts of the QNFs as $\alpha_0$ rises. This outcome suggests that the presence of quantum gravity effects leads to slower decay modes. We also briefly explore the impact of the charge parameter $q$. Our study indicates that the charge enhances the oscillations of the system. In addition, upon examining scenarios with higher angular quantum numbers ($l\geq 1$), we observe the disappearance of the non-monotonic behavior observed in the case of $l=0$. This disappearance may be attributed to the influence of the angular quantum number outweighing that of the quantum gravity effect. Further research is warranted to explore the universality of the non-monotonic behavior for $l=0$.

\acknowledgments

This work is supported by National Key R$\&$D Program of China (No. 2020YFC2201400), the Natural Science Foundation of China under Grants No. 12375055, No. 12347159 and No. 12305068, the Postgraduate Research $\&$ Practice Innovation Program of Jiangsu Province under Grant No. KYCX22$\_$3451, the Scientific Research Funding Project of the Education Department of Liaoning Province under Grant No. JYTQN2023090, and the Natural Science Foundation of Liaoning Province of China under Grant No. 2023-BSBA-229.

\bibliographystyle{style1}
\bibliography{Ref}
\end{document}